# *Improve accessibility for Low Vision and Blind people using Machine Learning and Computer Vision*

**Authors:**

*Jasur Shukurov*

**Contributors:**

*Aaqil Khoja*

*Giuseppe D'Ambrosio*

**Application Title:** Visual Companion

# Abstract:

With the ever-growing expansion of mobile technology worldwide, there is an increasing need for accommodation for those who are disabled. This project explores how machine learning and computer vision could be utilized to improve accessibility for people with visual impairments. There have been many attempts to develop various software that would improve accessibility in the day-to-day lives of blind people. However, applications on the market have low accuracy and only provide audio feedback. This project will concentrate on building a mobile application that helps blind people to orient in space by receiving audio and haptic feedback, e.g. vibrations, about their surroundings in real-time. The mobile application will have 3 main features. The initial feature is scanning text from the camera and reading it to a user. This feature can be used on paper with text, in the environment, and on road signs. The second feature is detecting objects around the user, and providing audio feedback about those objects. It also includes providing the description of the objects and their location, and giving haptic feedback if the user is too close to an object. The last feature is currency detection which provides a total amount of currency value to the user via the camera.

**Table of Contents**

## Introduction

Nearly 1 billion people, 15% of the world population, have some sort of permanent disability [1]. This world is not designed to accommodate people with disabilities. Thus there is a need to build software and hardware that will improve accessibility for those who experience challenges due to disability.

There are 14 main categories of disabilities [2]. Out of those 14, this paper will discuss the methods to improve accessibility for people who fall into categories such as blindness and visual impairment. The reasoning is that the world is getting more and more digitized. With the increase of digital technologies such as virtual reality, augmented reality, and non-fungible tokens (NFTs), there is progress towards more screen-based and vision-reliant software. If that benefits a wide majority of people [3], when it comes to visual impairment, it creates more barriers for blind people. Therefore, there is a greater need for modern technologies to be accessible to blind people.

According to the World Health Organization (WHO), there are roughly 253 million people in the world that suffer from some form of visual impairment, whether that is requiring eyewear or due to diseases such as glaucoma [4]. Of those 253 million, 36 million suffer from blindness due to diseases or were born with it. The WHO has estimated that it will only get worse by 2050. This paper will discuss how machine learning and computer vision could be utilized to improve accessibility for people with visual impairments. There have been many attempts to develop various software that would improve accessibility in the day-to-day lives of blind people. However, applications on the market have low accuracy, do not have user-friendly interfaces, or were not designed for blind people [17, 18, 19]. One of the applications is "ReCog" [17]. It provides users with the ability to discover objects. The limitations of this application are that users have to manually photograph the objects and train the object detection software. Another research, published in the *Journal of Springer Science-Business Media* in 2016 [18], demonstrates the use of machine learning to identify objects in real-time. However, this app offers only object detection features and provides only audio feedback.

The future digital world should consider the needs of underfunded groups in the creation and development of technologies. Technologies could be utilized to build tools to improve accessibility for people with disabilities. The primary methods were interviewing the target audience to identify the challenges they experience in their day-to-day life, and creating software using machine learning and computer vision to ease those challenges. Accompanying the paper will be mobile software that will help blind people orient in space by receiving audio and haptic feedback.

**Types of blindness:**

There are several different types of blindness and low vision [5]. According to The American Foundation for the Blind, individuals with low vision have "permanently reduced vision that cannot be corrected with regular glasses, contact lenses, medicine or surgery." Partial vision refers to the capacity to see only a portion of the visual field, or to have good central vision but poor peripheral vision. Total blindness is considered to be when a person sees total darkness. However, in most cases blind people can see to some extent; only 15% have total blindness [6]. During the initial interviews, it was determined that blind people do not like when they are being referred to as visually impaired. Therefore in this paper the term blind is being used to describe blindness, low vision and visual impairment.

# Methodology:

The first step was identifying what kind of challenges blind people experience in their day-to-day lives. For that purpose, 6 blind or low vision people with different social and economic backgrounds were interviewed. During the interview, the following questions were asked:

- What challenges do you have as a blind person ?
- What software do you use for accessibility ?
- What hardware do you use for accessibility ?
- What services do you use for accessibility ?
- What is lacking from that software ?
- What software would you like to use ?

The interview results showed that there are more than 25 types of challenges blind people experience while doing everyday things. Below are the main 8 challenges that most of the interviewees mentioned:

- Identifying the objects around them.
- Identifying currency.
- Identifying the amount of money.
- Reading signs and labels.
- Navigating from one place to another.
- Identifying the color of an object.
- Using software that does not have built-in accessibility for blind people.

According to the interviewees, there are multiple applications on the market which help with some of the challenges mentioned above. However, a big portion of those applications lacks several important aspects such as high accuracy, user-friendly interface accessible to blind people, text-to-speech interface, and audio responsive buttons.

Another important problem that was identified during the interview process was that users have to switch between applications to use different features. Most of the interviewees liked the idea of one application that will include several features and offer improved accessibility for blind users. For that purpose 3 core features were selected for an app:

- Scanning text and reading back to a user.
- Identifying objects in real time and describing them to a user.
- Scanning money and identifying the currency and amount of money.

**User Stories:**

| As a ... | I want to ... | So that |
| --- | --- | --- |
| Blind User | Be able to see the amount of money I have in my hand | I know how much money I have |
| Blind User | See what objects are in my vicinity | I do not run into them |
| Blind User | Be able to read any text that I have access to | I can understand the text |

# Requirements Discussion

*Assumptions:*

- The user is blind but able to hear.
- The user has a smartphone with access to the camera.
- No user authentication is needed for using the application.

*Functional Requirements:*

- This application will allow users to scan currency and relay back the type of currency, and the amount of money.
- This application will allow users to scan their surroundings to identify a selected object so that they may be able to locate it.
- This application will allow users to scan text and convert it into audio feedback.
- The application will utilize audio and haptic feedback to relay information/data back to the user.
- When the application is installed for the first time, it should display an introductory page.
- The system should be trained properly or adopt an Application Programming Interface (API - see glossary).

*Non-Functional Requirements:*

- The application should be user-friendly enough for a user to be familiar with navigating the application within the first five minutes of using it.
- The application will focus on the user experience.

- Software should be targeted at people who are blind.

- Software should be at least 70% accurate in identifying objects.

- Software UI should be responsive and adaptive across most mobile devices.

## Technical Discussion

**Development process:**

For the development process, the Agile methodology was used. The Agile development methodology is a series of approaches to software development focused on the use of iterative development, dynamic formation of requirements, and ensuring implementation is a result of constant interaction with a client and team members [7]. The development process was divided into 5 sprints of 2 weeks each. Each sprint is described below:

- Sprint 1 mostly consisted of the project setups and all of the preliminaries that were required for the project to move forward.
- Sprint 2 focused on the backend(see glossary) work for the project. The backend of the project comprised all component classes that were required for this project to work such as the functionality of the main page with the camera on the home screen, audio configuration with automatic voice, social interaction page, etc. After Sprint 2, the project prototype was shown to the clients. Based on feedback, it was determined that the application needed a very detailed introductory page with a user manual.
- Sprints 3 and 4 involved the use of the full front and back end development(see glossary); also known as full-stack development. Naturally, with this application that is for the visually impaired, it has little to no user interface(UI - see glossary) but focuses more on the user experience (UX - see glossary) aspect. Most of the UX is mentioned in the accessibility section. After the sprint 4 the feedback from the target audience was collected again. Several minor user experience issues were indicated which were addressed in the sprint 5.

- Sprint 5 focused on fixing the errors and testing the software system as a whole. Since this project is done in Android Studio which utilizes the Java programming language, an appropriate tool was chosen (JUnit) to facilitate writing tests for Java programs.

**Hardware choice:**

Until smartphones were introduced all products targeted for blind people were bulky and expensive [8]. That limited both consumers and businesses. For consumers, the equipment was too expensive. For business, it was not profitable. However, the era of smartphones changed this landscape. Most modern smartphones are equipped with high-quality cameras, Global Positioning System (GPS) trackers, sensors, and processors with high computational power. As a target hardware device for this application, we chose smartphones. There are several reasons for that choice. The primary reason is that smartphones are widely accessible. About 84% of the world population have smartphones, and this number will only increase [9]. The secondary reasons are that smartphones are easy to carry, applications could be easily installed, and it comes with a lot of pre-built accessibility features [21].

**Integrated development environment choice:**

Android Studio was chosen as an integrated development environment (IDE). It is an IDE produced by Google [10], which provides tools for creating applications on the Android Operating System (OS) platform. It provides an interactive and flexible environment for mobile application development. The primary programming language used was Java. Other programming languages such as Kotlin and Python were used for object detection.

**API choice:**

Machine learning is a subfield of artificial intelligence that deals with computer algorithms that learn from training data without being explicitly programmed. Computer vision is a field of artificial intelligence related to image and video analysis. It includes a set of methods that give the computer the ability to see and extract information from what it sees. Utilizing those two tools, it is possible to develop an application that will to some extent replicate human vision.

Training models for machine learning and computer vision require enormous computational power. However, there are available computer vision models that have already been trained and have high accuracy. One of those tools is Google Vision API [11]. It is a very popular API when it comes to computer vision. Another API was used to detect objects, Tensorflow object detection API [12], an open-source framework, thus facilitating integration with the rest of the project's code.

**Improved Accessibility features:**

The first step is to provide the user with information about their vicinity. The object detection API returns a text description of the object seen by the smartphone's camera. Since audio feedback is desired, this text needs to be converted to speech. The conversion to audio is accomplished by Android's software library dedicated to text-to-speech. Furthermore, a text to speech feature has been implemented in every page of the application, so when the user clicks on any button, it reads back the button's functionality. To use the button, the user has to press and hold it. Most of the smartphones are equipped with a back button. However, those buttons are tiny and very hard to locate for a blind person. Therefore, swipe features have been added. In order to proceed to the previous page, a user has to swipe down to up.

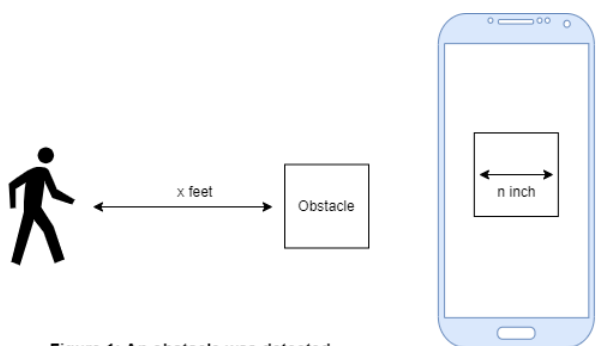

In addition, a touch interface (haptics) is used as secondary feedback. When a user is in a relatively normal distance to a specific object (~5-10ft), the app starts off with medium haptics, meaning it gives some vibration. Eventually, the haptics increase the response towards the user as they approach an absolute zero, which is when the user is right next to an object. To determine the initial distance to an object, the OpenCV library is used. OpenCV is an open source computer vision and machine learning software library [20]. Once the initial distance is determined, it is recorded and stored in the local database. As the user gets closer to the object, the size of the object within the camera vicinity increases. Based on how much the size of the object has increased compared to the initial size, the distance between the object and the user is measured.

Besides, the haptics are being used when the user navigates from page to page. When a user opens a new page, it gives one long vibration, meaning that the user opened a new page. When a user goes back to the previous page, it gives 3 short vibrations to notify a user that they have moved to the previous page.

The front-end work is focused on the user experience. Therefore the decision was made for simplicity of user interface without the need for visually-appealing elements. The main priority is functionality.

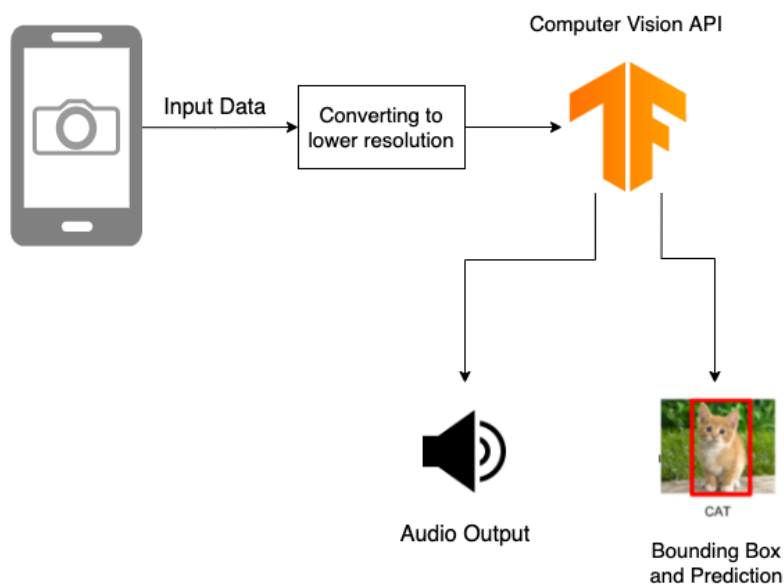

**Input Requirements and Quality:**

The quality of the input is crucial when it comes to accuracy and speed of detection. There is always a trade off between accuracy and speed. Higher quality images will provide more accurate predictions, but at the cost of sacrificing speed. This is due to the fact that high quality

14images have higher density of pixels which require more processing time. Therefore, input video and images are first converted into lower resolution and sent to a machine learning algorithm, which then determines the result.

**High-Level Design:**

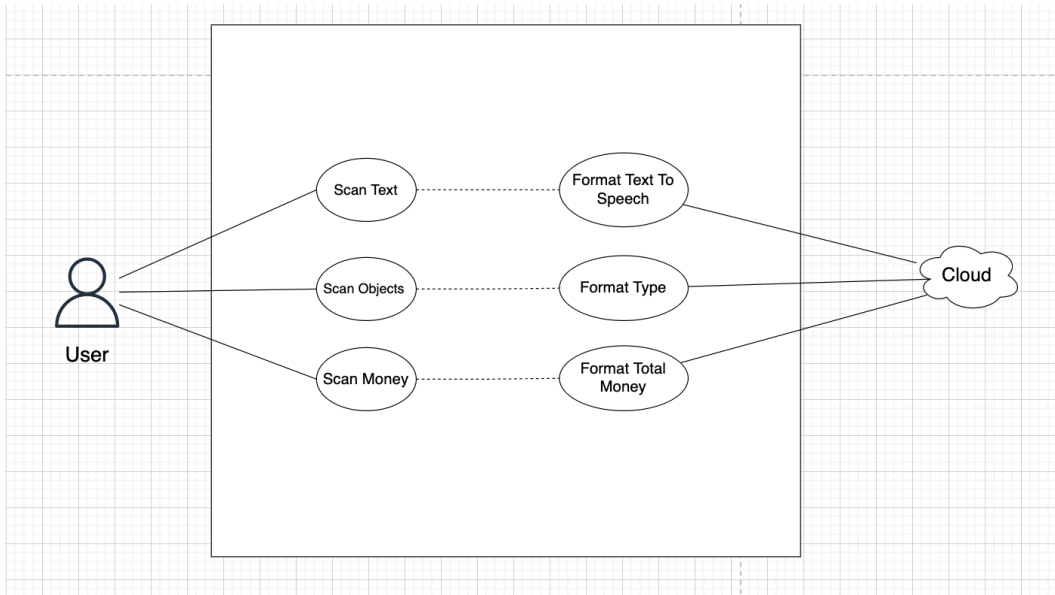

*Figure #1. Use case Diagram, Visual Companion App*

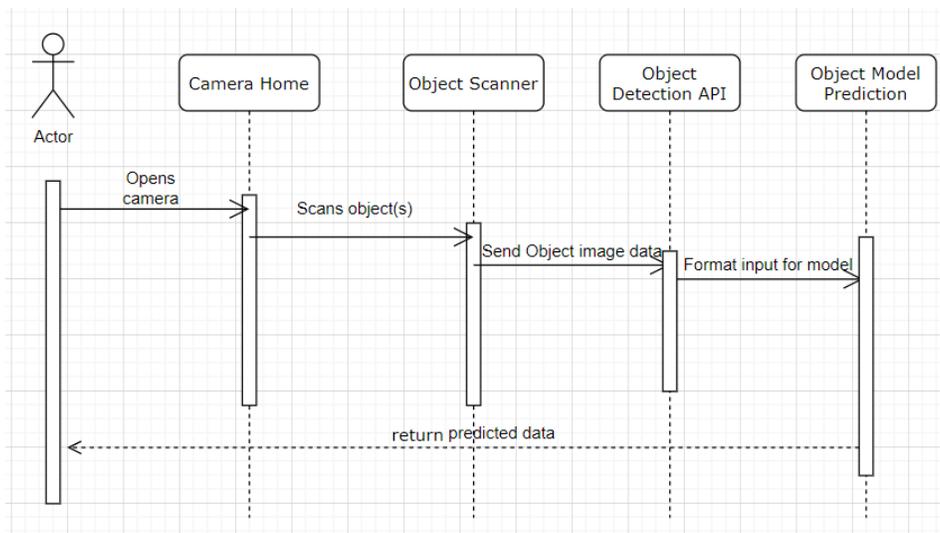

*Figure #2. Sequence Diagram, Visual Companion App*

**User Manual:**

The app is simple to use. When a user installs the application and launches it for the first time, it proceeds to the introduction page. The introduction page explains how to use the app, providing the instructions over audio. Text instruction is also provided.

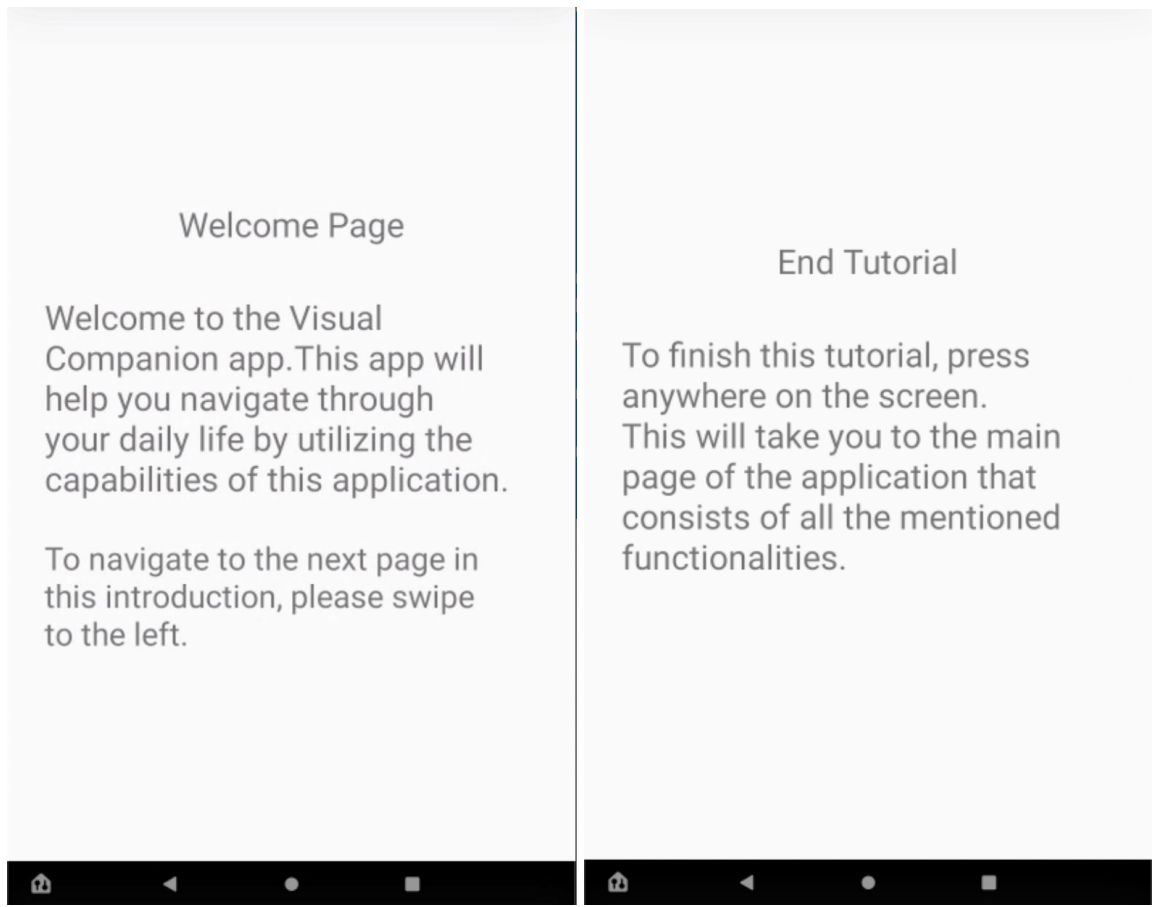

*Figure #3. The first and the last pages of the Introduction Activity, Visual Companion App. (See Appendix A for all screenshots)*

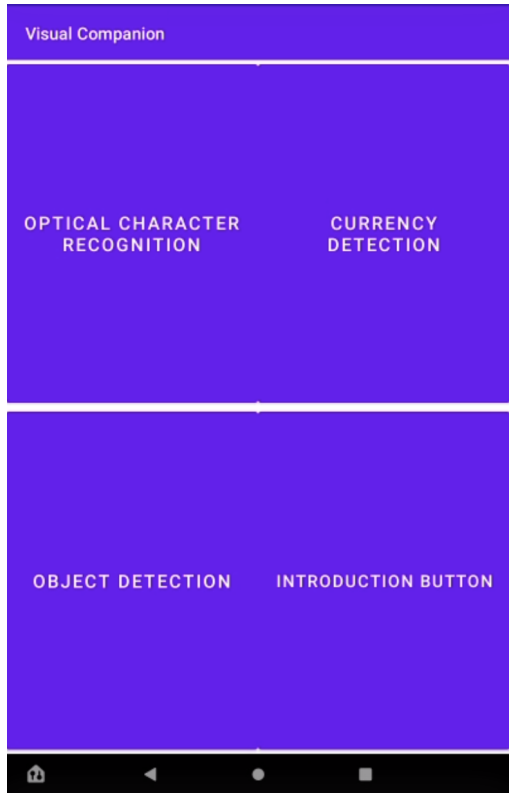

*Figure #4. Front Page, Visual Companion App*

Once the introduction is finished, the user will proceed to the front page. The front page features 4 buttons. One click on the button tells the functionality of it, and a long press will take the user to the page associated with the selected functionality.

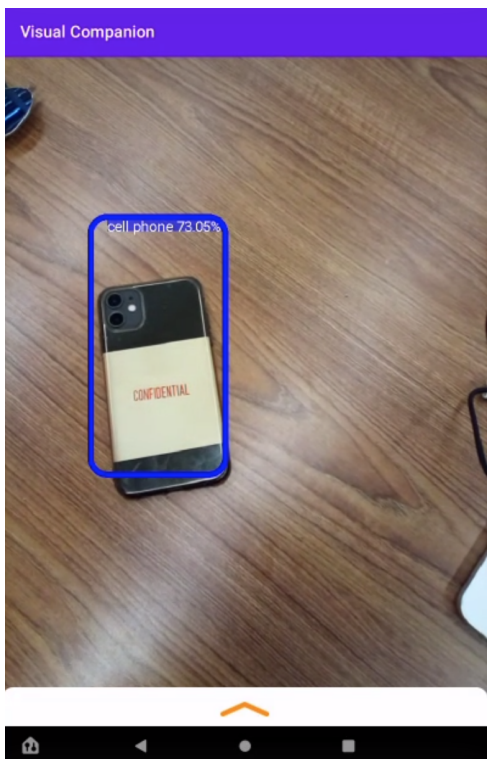

*Figure #5. Object Detection Features, Visual Companion App*

The first feature is object detection. To use this feature, the user has to press and hold the object detection button on the front page. The application will detect objects in the immediate vicinity of the camera and inform the user about these objects and their location.

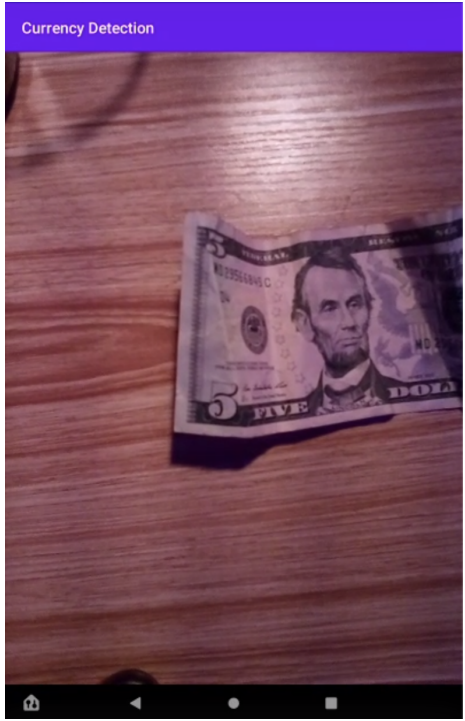

*Figure #6. Currency Detection Features, Visual Companion   App*

In the currency detection page, users can just direct the phone towards money,  and press anywhere on the screen. The app will detect the currency and tell the amount of money.

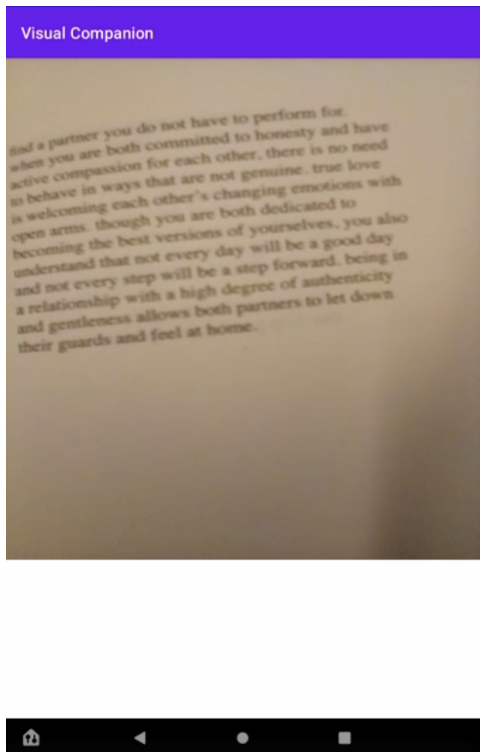

*Figure #7. Optical Character Recognition, Visual Companion App*

To use this feature the user has to direct the phone toward any text, and press anywhere on the screen. It will automatically detect the text, and read it back to the user.

## Summary/Conclusion

The interviews and research demonstrated a need and demand for the applications that improve the accessibility for blind people. According to the conducted interviews there are more than 25 challenges blind people experience while performing daily activities. Thus, this paper discusses building an application that helps to ease challenges such as:

- Object Detection
- Currency Detection
- Reading any text.

In accordance with best practices, using the methodology of agile software development, the software was developed in stages, or sprints, with feedback from the blind community sought after each sprint, to ensure that the app effectively meets their needs.

## Glossary of Terms:

| Term | Definition |
|---|---|
| API | acronym for Application Programming Interface, an interface that allows our software to talk to another application through requests in order to use its services. |
| UI | acronym for User Interface, essentially what the user will be interacting with in order to use the app. |
| UX | the concept of the entire experience that a user receives when interacting with an application. |
| Functional Requirement | a list of services that must be performed by the system. |
| Non-functional Requirements | system qualities, i.e. desirable attributes of the overall system such as reliability, performance, maintainability, scalability, and usability. |
| User story | describes a software feature from the perspective of the user, i.e. how the feature provides value to the user. |
| Frontend | the client part of the product, the interface with which the user interacts. |
| Backend | the internal part of the product, which is located on the server and is hidden from users. |

## Appendix A: User Manual

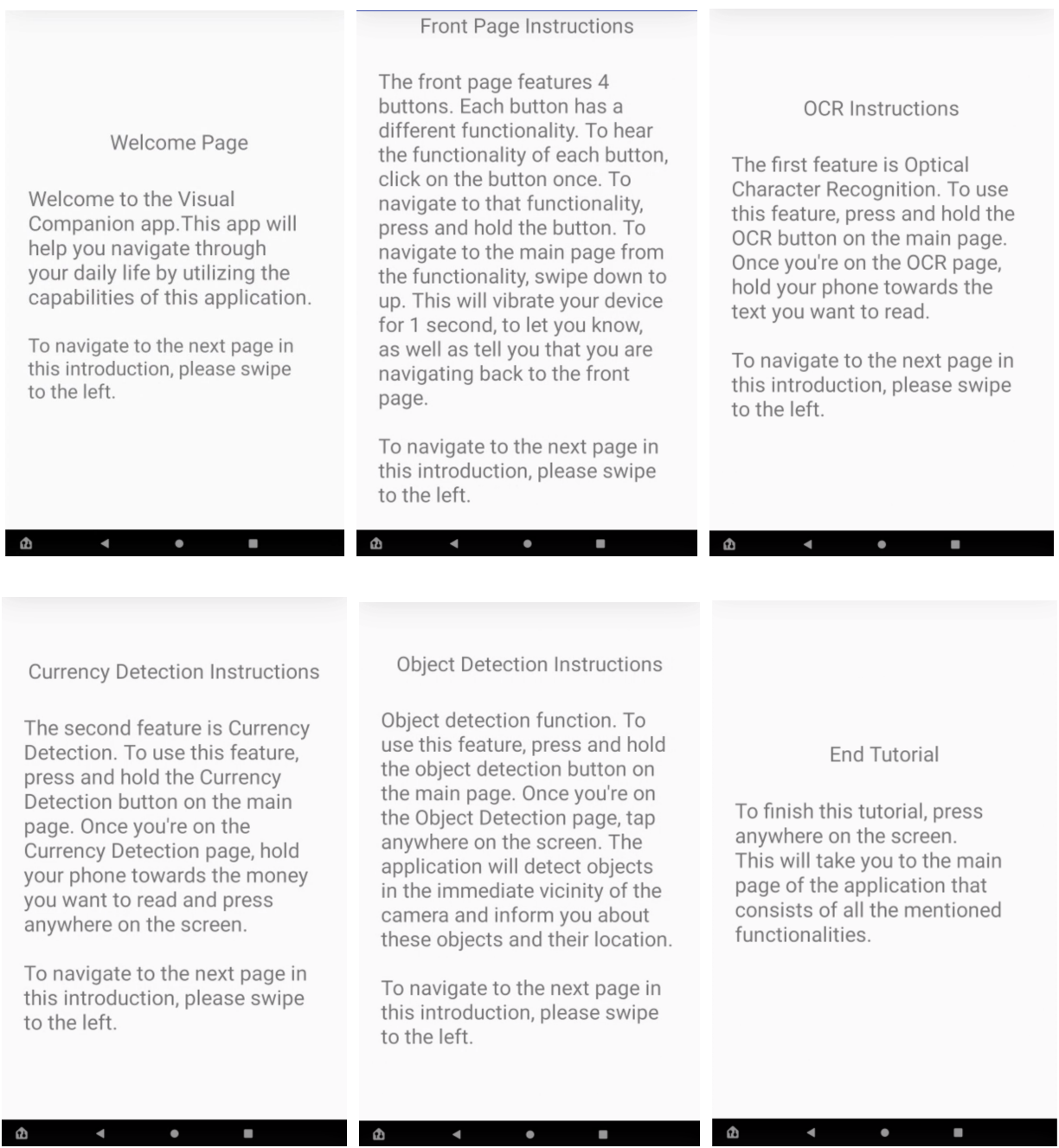

**References:**

---